\documentclass[twocolumn,usenatbib]{mnras}

\usepackage{graphicx}

\newcounter{parentequation}
\def\beglet{
  \addtocounter{equation}{1}%
  \setcounter{parentequation}{\value{equation}}%
  \setcounter{equation}{0}%
  \def\theequation{\arabic{parentequation}\alph{equation}}%
  \ignorespaces
}
\def\endlet{
  \setcounter{equation}{\value{parentequation}}
  \def\theequation{\arabic{equation}}%
}
\def\pmb#1{\setbox0=\hbox{#1}%
    \kern-.025em\copy0\kern-\wd0
    \kern.05em\copy0\kern-\wd0
    \kern-.025em\raise.0433em\box0}

\def\ltsima{$\; \buildrel < \over \sim \;$}
\def\gtsima{$\; \buildrel > \over \sim \;$}
\def\simlt{\lower.5ex\hbox{\ltsima}}
\def\simgt{\lower.5ex\hbox{\gtsima}}
\def\p2Y{\;_2Y}
\def\m2Y{\;_{-2}Y}

\def\mk2{\mu {\rm K}^2}
\def\Planck{\it Planck\rm}
\def\Plancks{\it Planck \rm}
\def\LCDM{$\Lambda$CDM}

\def\pmb#1{\setbox0=\hbox{#1}%
     \kern-.025em\copy0\kern-\wd0
     \kern.05em\copy0\kern-\wd0
     \kern-.025em\raise.0433em\box0}

\begin{document}

\title[Statistical Inconsistencies in KiDS]{Statistical Inconsistencies in the KiDS-450 Dataset}

\author[George Efstathiou and Pablo Lemos]{George Efstathiou and Pablo Lemos \\
 Kavli Institute for Cosmology Cambridge and 
Institute of Astronomy, Madingley Road, Cambridge, CB3 OHA.}

\maketitle

\begin{abstract} The Kilo-Degree Survey (KiDS) has been used in several recent papers to infer
constraints on the amplitude of the matter power spectrum and matter
density at low redshift.  Some of these analyses have claimed tension
with the \Planck\ \LCDM\ cosmology at the $\sim 2-3\sigma$ level,
perhaps indicative of new physics.  However, \Plancks is consistent
with other low redshift probes of the matter power spectrum such as
redshift space distortions and the combined galaxy-mass and
galaxy-galaxy power spectra.  Here we perform consistency tests of the
KiDS data, finding internal tensions for various cuts of the data at
$\sim 2.2 - 3.5\sigma$ significance.  Until these internal tensions
are understood, we argue that it is premature to claim evidence for
new physics from KiDS. We review the consistency between KiDS and other
weak lensing measurements of $S_8$, highlighting the importance of
intrinsic alignments for precision cosmology.

\vskip 0.3 truein

\end{abstract}

\section{Introduction}

Precision observations of the cosmic microwave background radiation
(CMB) by the Planck satellite \citep[][hereafter P16]{Planckparams14,
  Planckparams16} and  other experiments \citep{WMAP:13,ACT13, SPT13}
have shown that the \LCDM\ cosmology, with nearly scale invariant, adiabatic,
Gaussian initial perturbations, provides an excellent description of
our Universe. Measurements of weak lensing of the CMB
\citep{Plancklensing16} show further that the \LCDM\ model remains a
good description of the Universe down to a redshift of $z\sim 2$,
where the CMB lensing kernel peaks.

It is, nevertheless, important to test the model at lower redshifts,
particularly at redshifts $z \simlt 1$ when the Universe becomes
dominated by dark energy. Deviations from the \LCDM\ model at low
redshift could potentially reveal evidence for dynamical dark energy
or modifications to General Relativity \citep[see][for a
review]{Amendola:16}.

Weak galaxy lensing is an  important probe of the matter
power spectrum at low redshifts
\citep{Blandford:91,Miralda-Escude:91,Kaiser:92}. Several ambitious
deep imaging projects have reported results recently. These include
the Canada France-Hawaii Telescope Lensing Survey \citep[CFHTLenS,][]{Heymans:12, Heymans:13,Joudaki:17a},  Deep Lens Survey \citep[DLS,][]{Jee:16}, Dark Energy Survey \citep[DES,][]{Abbott:16,Troxel:17, DES:17} and 
Kilo Degree Survey \citep[KiDs,][]{Hildebrandt:17, Kohlinger:17}. Weak lensing analysis of these
surveys can be used to constrain the parameter combination\footnote{Where
$\sigma_8$ is the present day linear theory root-mean-square amplitude of the matter fluctuation spectrum
averaged in spheres of radius $8 \; h^{-1} {\rm Mpc}$, $\Omega_m$ is the present day matter density in units 
of the critical density $\rho_c$ and $h$ is the Hubble constant in units of $100 \ {\rm km}\ {\rm s}^{-1}
{\rm Mpc}^{-1}$.}
$S_8 = \sigma_8(\Omega_m/0.3)^{0.5}$, which can be compared to the \Planck\ value from P16\footnote{Unless stated otherwise, we quote $\pm 1\sigma$ errors on parameters.},
$S_8 = 0.825 \pm 0.016$ derived from the Planck temperature power spectrum, low multipole polarization and Planck lensing (TT+lowTEB+lensing, in the notation of P16). However, the weak 
galaxy lensing results span a range of values. The reanalysis of CFHTlenS by \cite{Joudaki:17a} 
finds $S_8 = 0.732^{+0.029}_{-0.031}$; \cite{Jee:16} find $S_8 = 0.818^{+0.034}_{-0.026}$ from DLS;
\cite{Abbott:16} find $S_8 = 0.81 \pm 0.06$ from the DES Science Verification data; \cite{Hildebrandt:17} (hereafter H17) find $S_8 = 0.745 \pm 0.039$ from a  tomographic correlation function analysis
of KiDs while \cite{Kohlinger:17} (hereafter K17) find $S_8 = 0.651 \pm 0.058$ from a tomographic power spectrum 
analysis of KiDs. The  DES Year 1 weak lensing analysis\footnote{DES Year 1 results \citep{Troxel:17, DES:17} appeared after the submission of this paper and so will not be discussed in detail.} \citep{Troxel:17} gives $S_8 =0.789^{+0.024}_{-0.026}$. Some of these values are in tension with \Planck. For example, 
H17 find a $2.3\sigma$ discrepancy between KiDs and \Planck, while K17 find a $3.2\sigma$ discrepancy. However, 
the results  from these different surveys do not agree particularly well with each other (even when using the same shear catalogue), showing differences in the value of $S_8$ 
at the   $\sim 2-2.5 \sigma$ level. 

A statistically significant tension between the \Planck\ \LCDM\ cosmology and weak galaxy 
lensing could have important consequences for fundamental physics \citep[e.g.][]{Joudaki:16}.
But how seriously should we take the weak lensing results? A minimal requirement 
is that a cosmic shear data set should be internally self-consistent. The main purpose of
this paper is to show that this does not seem to be the case with KiDS.

Before we begin, we make a few remarks concerning cosmic shear analysis. Most analyses involve
estimation of correlation functions $\xi_+$ and $\xi_-$ as a function of relative angular separation $\theta$, or of the cosmic shear E-mode 
power spectrum $P_\kappa(\ell)$ as a function of multipole $\ell$. These are related by 
\begin{equation}
\xi_{\pm} = {1 \over 2 \pi} \int d\ell \ell P_\kappa(\ell)J_{0,4} (\ell \theta).
\end{equation}
For a cross-power spectrum between redshift bins $i$ and $j$, the shear power spectrum is
related to the non-linear matter power spectrum $P_\delta$ by 
\begin{equation}
 P^{ij}_\kappa(\ell) = \int_0^{\chi_H} d\chi {q_i(\chi) q_j(\chi) \over [f_K(\chi)]^2} P_\delta
\left ( {(\ell+1/2) \over f_K(\chi)}, \chi \right ),
\end{equation}
where (following the notation of H17) $\chi$ is the comoving radial distance, 
$f_K(\chi)$ is the comoving angular diameter distance to distance $\chi$,
and $q_i(\chi$) is the lensing efficiency for tomographic redshift bin $i$:
\begin{equation}
 q_i(\chi) = {3 H_0^2 \Omega_m \over 2 c^2} {f_K(\chi) \over a(\chi)} \int_\chi^{\chi_H} 
d\chi^\prime n_i(\chi^\prime) {f_K(\chi^\prime - \chi) \over f_K(\chi^\prime)},
\end{equation}
where  $\chi_H$ is the 
comoving Hubble distance and $n_i(\chi)$ is the effective (weighted) number density galaxies in redshift bin $i$ normalized
so that $\int n_i (\chi) d\chi = 1$. Even if the image analysis is assumed to be 
free of systematic errors and biases, inferences on cosmology require an accurate model
of the redshift distribution $n_i(\chi)$, which in turn requires accurate calibration of the
photometric redshifts used to define the redshift bin $i$. A key test of the accuracy of the
photometric redshift calibrations would be to demonstrate consistency between 
distinct cross-correlations $i, j$. However, this is not straightforward because of intrinsic
ellipticity 
alignments  between neighbouring galaxies (II term) and between gravitation shear and intrinsic
shear (GI term).  The power spectra\footnote{Neglecting B-modes.} of these  terms are
 usually modelled as
\beglet
\begin{eqnarray}
 & & {\hskip -0.49 truein}
 P^{ij}_{II}(\ell) =\int_0^{\chi_H} d\chi F^2(z) {n_i(\chi) n_j(\chi) \over [f_K(\chi)]^2} P_\delta
\left ( {(\ell+1/2) \over f_K(\chi)}, \chi \right ),   \label{IA1} \\
& & {\hskip -0.49 truein}  P^{ij}_{GI}(\ell) =\int_0^{\chi_H} d\chi F(z) {(q_i(\chi) n_j(\chi) + n_i(\chi)q_j(\chi)) \over [f_K(\chi)]^2}      \nonumber \\
 & & \qquad \times P_\delta\left ( {(\ell+1/2) \over f_K(\chi)}, \chi \right ), \label{IA2}
\end{eqnarray}
\citep{Hirata:04, Bridle:07}.
In these equations, 
\begin{equation}
  F(z) = -A_{IA} C\rho_c {\Omega_m \over D(z)},  \label{IA3}
\end{equation}
\endlet
where $D(z)$ is the linear growth rate of perturbations normalized to unity at the present day,
and C is a normalizing constant, usually chosen to be $C = 5 \times 10^{-14} h^{-2} M_\odot^{-1} {\rm Mpc}^3$. With this choice, the intrinsic alignment amplitude is expected to be of order unity (and
positive if intrinsic ellipticities are aligned with the stretching axis of the tidal field).  This
model of intrinsic alignments is heuristic and simplified (see \cite{Blazek:17} for a more complex alignment  model). 
Even in the context of this 
model, the intrinsic alignment amplitude may vary with redshift, luminosity, and galaxy type. For current weak lensing surveys, intrinsic alignments are not benign. The contributions of equs. \ref{IA1}
and \ref{IA2} are comparable to any claimed tensions between the \Plancks value of $S_8$ 
and those inferred from cosmic shear surveys (with positive $A_{IA}$ tending to raise the value of $S_8$ and
negative values lowering $S_8$). How can we test the intrinsic alignment model?  The conventional
solution is to introduce additional nuisance parameters to characterize  uncertainties in the intrinsic alignment model \citep[e.g.][]{Kirk:12}, relying on the redshift dependence of the measured signals to disentangle true cosmic shear from intrinsic alignments. This, of course, requires accurate
knowledge of the redshift distributions and their errors. 

Current cosmic shear data is still relatively sparse, with a small number of measurements in coarse
redshift bins. The number of internal consistency checks of the data and the various components of the model (including nuisance parameters) is therefore 
limited\footnote{The situation is very different to the CMB, where there is a large amount
of information to separate a high amplitude frequency independent cosmological
signal with a distinctive power spectrum from low amplitude foregrounds with smooth
power spectra.}. In Section 2, we perform consistency tests of the KiDS data from 
H17. In Section 3 we compare the KiDS results with \Plancks and measurements of redshift space distortions and rich cluster abundances, which provide
independent measures of the amplitude of the matter fluctuations at similar redshifts to those of the KiDS galaxies. Section 4 
compares the results from  various weak lensing analyses. Our main conclusions are
presented in Section 5.


\section{Tests of the  KiDS data}

\begin{figure}
	\centering
	\includegraphics[width=73mm, angle=0]{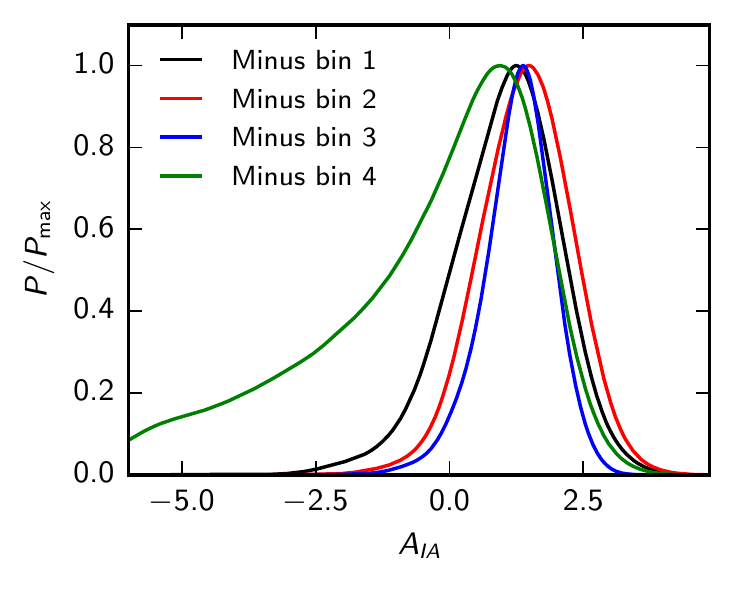}	\includegraphics[width=78mm, angle=0]{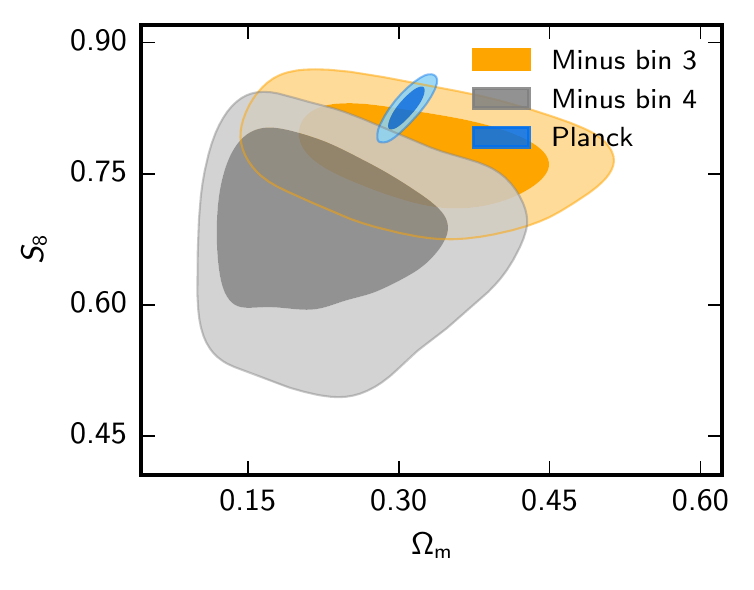}
	\caption{The upper panel shows the posteriors for the intrinsic alignment parameter $A_{IA}$
(equ. \ref{IA3}) as we remove all cross-correlations involving a particular redshift bin. The lower
panel shows the 68 and 95\% constraints on $S_8$ for the data minus redshift bin 3 (orange) and minus redshift bin 4 (grey). The blue contours show the \Plancks constraints from the TT+lowTEB+lensing
data combination as given in P16.}
	\label{fig:ia+s8}

\end{figure}

We use the KiDS cross-correlation measurements of $\xi_+$ and $\xi_-$
in four tomographic redshift bins as reported by H17 together with the
associated {\tt CosmoMC} likelihood module and covariance
matrix\footnote{Downloaded from http://kids.strw.leidenuniv.nl.}.  For
reference, the four redshift bins span the following ranges in
photometric redshift $z_B$: $0.1 < z_B \le 0.3$ (bin 1), $0.3 < z_B \le
0.5$ (bin 2), $0.5 < z_B \le 0.7$ (bin 3), $0.7 < z_B \le 0.9$ (bin 4).
We used the same angular ranges, photometric redshift calibrations and
errors, nuisance parameters and priors as in `fiducial' analysis in H17
(first entry in their Table 4) and verified that we recovered the
identical best-fit $\chi^2$ ($162.8$) and constraint on $S_8$ ($S_8 =
0.745 \pm 0.039$). We then removed all cross-correlations involving one
of the photometric redshift bins. The results are summarized in Table
\ref{tab:kids_cond_all} and in Fig. \ref{fig:ia+s8}.

\begin{table}
{	\centering
	\caption{Conditional $\chi^2$ tests removing photometric redshift bins}
	\label{tab:kids_cond_all}
	\begin{tabular}{ccccc} 
		\hline
        ${\bf y}^D$       &  $S_8$ &  $A_{IA}$  &  $\chi^2_{\rm cond}$  &  $N_{\sigma_{\rm cond}}$ \\ 
 minus  z-bin 1    & $0.745 \pm 0.040$ & $1.14 \pm 0.85$ & $61.0$ (52) &    $0.89$\\
 minus  z-bin 2    & $0.754 \pm 0.042$ & $1.24 \pm 0.80$ & $66.3$ (52) &    $1.40$\\
 minus  z-bin 3    & $0.771 \pm 0.039$ & $1.25 \pm 0.57$ & $78.2$ (52) &    $2.60$\\
 minus  z-bin 4    & $0.684 \pm 0.071$ & $-0.1 \pm 1.7$  & $87.9$ (52) &     $3.52$\\
 minus  $\xi_-$    & $0.778 \pm 0.040$ & $1.10 \pm 0.73$ & $89.7$ (60) &     $2.71$\\
 minus  $\xi_+$    & $0.705 \pm 0.048$ & $0.92 \pm 0.97$ & $84.1$ (70) &    $1.20$\\
		\hline
              \end{tabular}}
            {\small Notes: The first column defines the portion of the data vector  (${\bf y}^D$) used 
to fit the model. The second and third columns
give the marginalised mean values of $S_8$, $A_{IA}$ and their $1\sigma$ errors. The fourth column gives the conditional $\chi^2_{\rm cond}$,   
as defined in equ. \ref{cond3}, for the rest of data vector, ${\bf x}^D$. The numbers in parentheses list the length, $N_{\rm x}$, of the vector ${\bf x}^D$. 
 The fifth column gives the number of standard deviations by which  $ \chi^2_{\rm cond}$ differs from $N_{\rm x}$, $N_{\sigma_{\rm cond}} = 
              ( \chi^2_{\rm cond} - N_{\rm x})/\sqrt{2N_{\rm x}}$.}

\end{table}

 The first point to note is that the intrinsic alignment amplitude is
 reasonably stable to the removal of photometric redshift bins.  All
 of the posteriors shown in Fig. \ref{fig:ia+s8} are consistent with
 the intrinsic alignment solution from the full dataset ($A_{IA} =
 1.10^{+0.68}_{-0.54}$). However, it is also clear that redshift bin 4
 carries a high weight in fixing $A_{IA}$. With redshift bin 4
 removed, the posterior distribution develops a long tail to negative
 values that is cut-off by the lower end of the $A_{IA}$ prior
 (uniform between $-6 < A_{IA} < 6$). As a consequence of this long
 tail, the best fit value of $S_8$ with bin 4 removed is driven to
 lower values and its error increases substantially compared to the
 full sample (lower panel of Fig. \ref{fig:ia+s8} and Table
 \ref{tab:kids_cond_all}). Redshift bin 4 is therefore critical in
 pinning down the intrinsic alignment solution and reducing the error
 on $S_8$.

If redshift bin 3 is removed, $S_8$ rises and the constraints in the
$S_8-\Omega_m$ plane become compatible with \Plancks
(Fig. \ref{fig:ia+s8}).  This is not unexpected, because one can see
from Fig. 5 of H17 that the best-fit fiducial model tends to sit high
for all cross-spectra involving tomographic redshift bin 3
(particularly for $\xi_{-}$). With redshift bin 3 removed, there is
substantial overlap in the posteriors in the $S_8-\Omega_m$ plane with
those from the full sample and with the other subsets of the data
summarized in Table \ref{tab:kids_cond_all}. However, these various
estimates of $S_8$ are highly correlated since they share common
data. Are the parameter shifts seen in these subsets statistically
reasonable? We turn to this question next.

\begin{figure}
	\centering
	\includegraphics[width=83mm, angle=0]{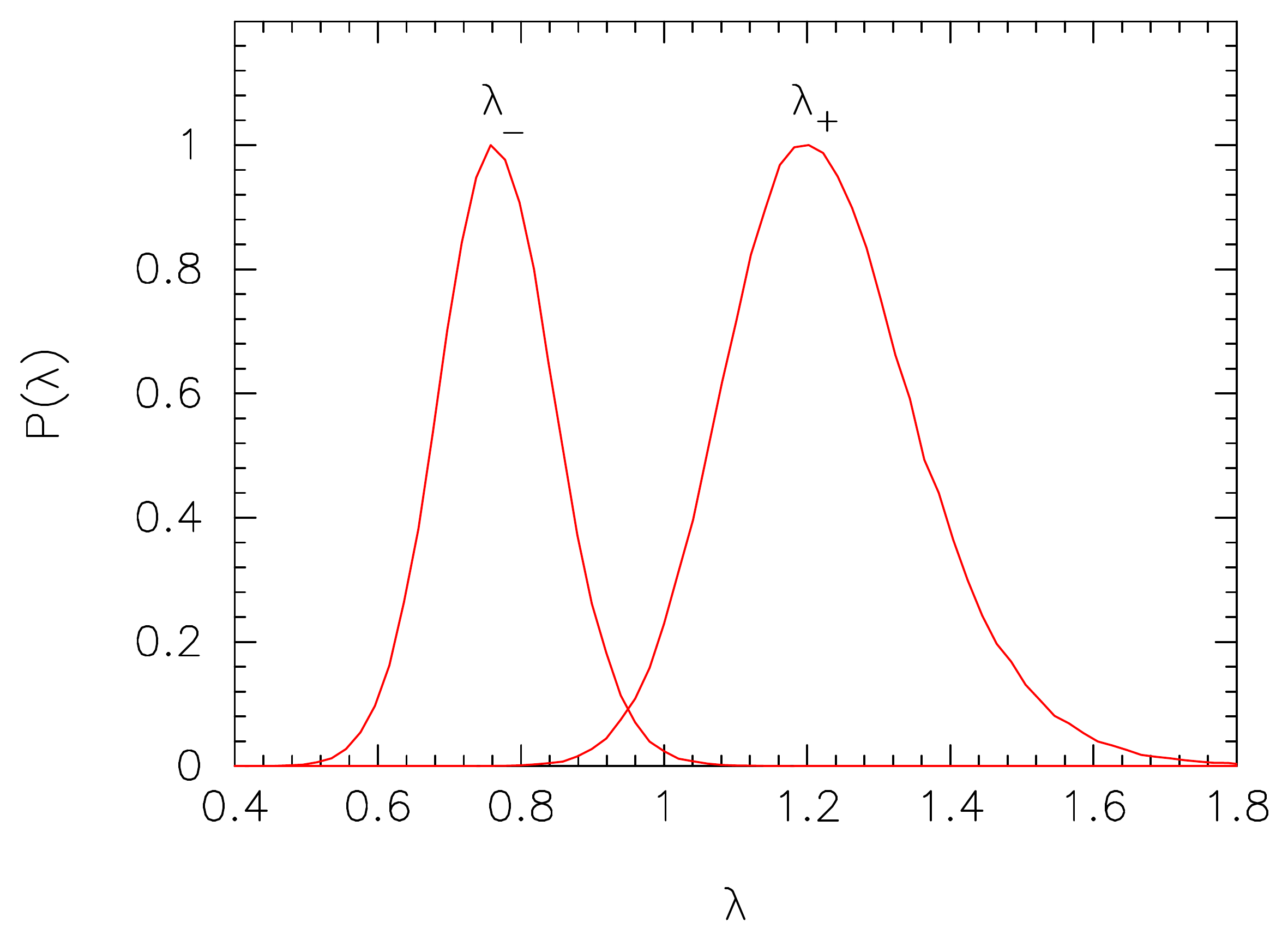} \\
	\includegraphics[width=83mm, angle=0]{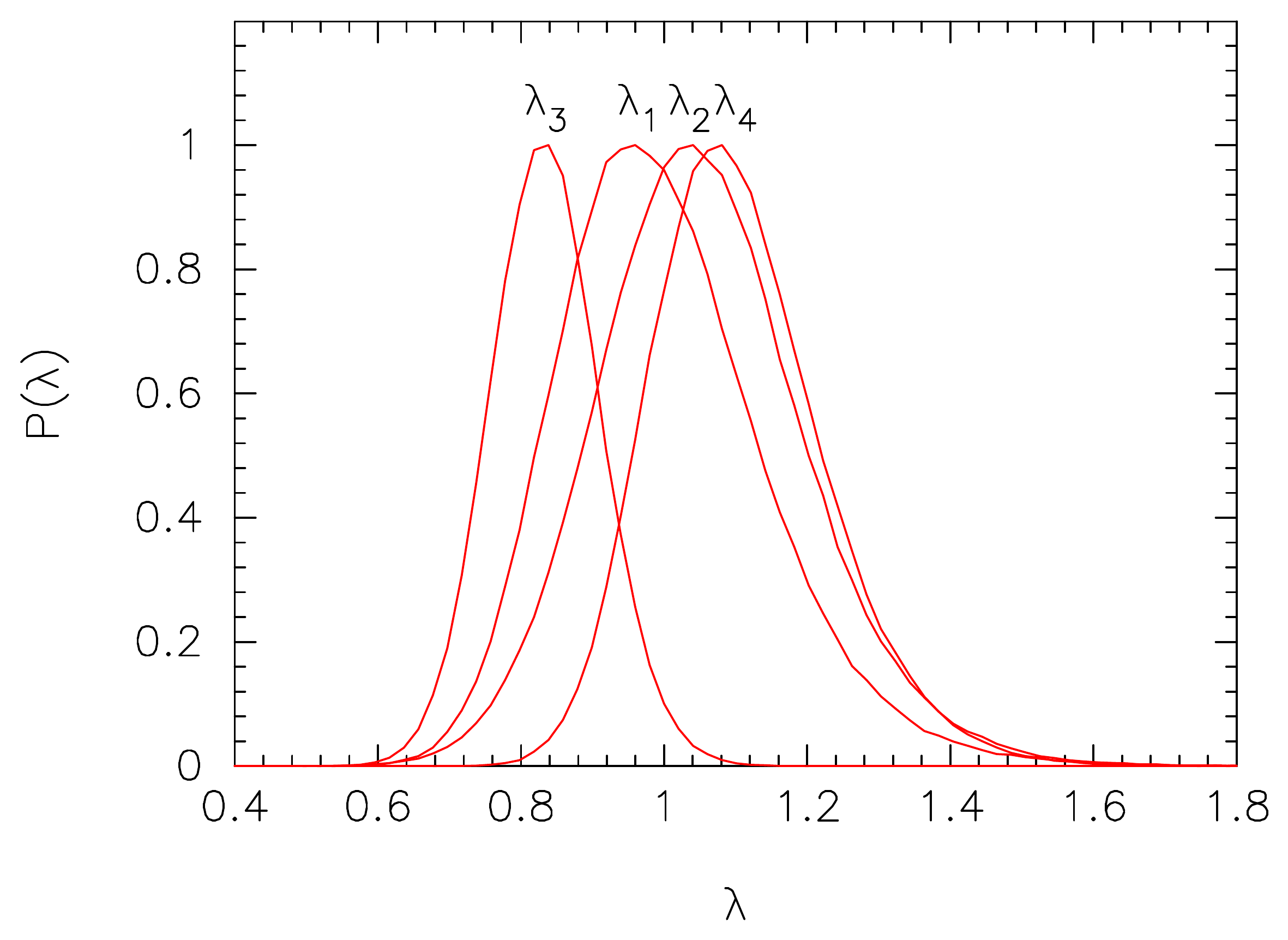}
\vskip -0.07 truein
	\caption{Posterior distributions of the parameter $\lambda$ defined in Equ. \ref{Eq:lambda}. 
The upper figure shows the distributions if the model parameters are fitted to $\xi_+$ (denoted $\lambda_-$) 
and to $\xi_-$ (denoted $\lambda_+$). The lower figure shows the posterior distributions of $\lambda$ for
partitions of the data in which all cross-correlations involving a particular tomographic redshift bin are removed from the fit to the theoretical model (e.g. $\lambda_3$, corresponds to a theoretical model fitted to all cross-correlations that do not
involve tomographic redshift bin $3$).}
	\label{fig:lambda}

\end{figure}

We can perform a more elaborate statistical consistency test by dividing the data
vector into two components:
\begin{equation}
      {\bf z}^D = ({\bf x}^D, {\bf y}^D).
\end{equation}
We can then fit ${\bf y}^D$ to a model (including nuisance parameters), $\hat {\bf y}$. The model parameters also make a 
theory prediction for the data partition ${\bf x}^D$, which we denote ${\hat {\bf x}}$. We can then write the theory vector
for ${\bf z}^D$ as 
\begin{equation}
      {\hat {\bf z}} = (\lambda{\hat {\bf x}}, {\hat {\bf y}}),  \label{Eq:lambda}
\end{equation}
introducing a new parameter $\lambda$. Evidently, if the data partitions and model are consistent, the new parameter $\lambda$ should be consistent with
unity. The tests described in this Section are all based on the \LCDM\ model, {\it but with a free amplitude}. Since cosmic shear measurements have very limited ability to
fix shape parameters, and the data cuts that we apply cover similar redshift ranges, 
it seems reasonable to interpret differences in $\lambda$ as indicative of systematic errors in the data. To recap, we run MCMC chains to determine the model
parameters from a data partition ${\bf y}^D$ and determine a single amplitude parameter $\lambda$ by fitting to the rest of the data ${\bf x}_D$.  The posterior
distributions of $\lambda$ for the data cuts of Table \ref{tab:kids_cond_all} are shown in Fig.  \ref{fig:lambda}.

The upper plot in Fig. \ref{fig:lambda}  compares the amplitudes $\lambda_-$ (fitting the model parameters to $\xi_+$ ) and $\lambda_+$ (fitting the model parameters to $\xi_-$). This agrees with the visual impression given by Fig. 5 of H17, namely that $\xi_-$ wants a low amplitude while $\xi_+$ prefers a high amplitude.   Integrating these distributions, 
\beglet
\begin{eqnarray}
\int_0^{1} P(\lambda_-) d\lambda_- = 2.9\times 10^{-3},  \qquad  \qquad\\
\int_1^\infty P(\lambda_+) d\lambda_+ =  4.2 \times 10^{-2}. \qquad \qquad
\end{eqnarray}
A value of $\lambda=1$ therefore lies in the tails of both posterior distributions.  These
results show that $\xi_-$ sits about $2.8\sigma$ low compared to the best fit
\LCDM\ cosmology determined from $\xi_+$. 

The lower plot in Fig. \ref{fig:lambda} 
tests consistency between photometric redshift bins including both $\xi_+$ and $\xi_-$ in the fits. 
The parameters $\lambda_i$ (with $i$ running from $1-4$) are computed for data partitions in which ${\bf y}^D$  excludes
all cross-correlations involving photometric redshift bin $i$. In this test, photometric redshift bin 3 is an outlier
with 
\begin{equation}
\int_0^{1} P(\lambda_3) d\lambda_3 = 1.3\times 10^{-2},  \qquad  \qquad
\end{equation}
\endlet
suggesting that the data involving photometric redshift bin 3 is inconsistent with the rest of the data at about the $2.2\sigma$ level. Again, this accords with the visual impression from Fig. 5 of H17, which shows that cross-correlations in both
$\xi_+$ and $\xi_-$ involving photometric redshift bin 3 tend to lie below their best fit model.

\begin{figure}
	\centering
	\includegraphics[width=85mm, angle=0]{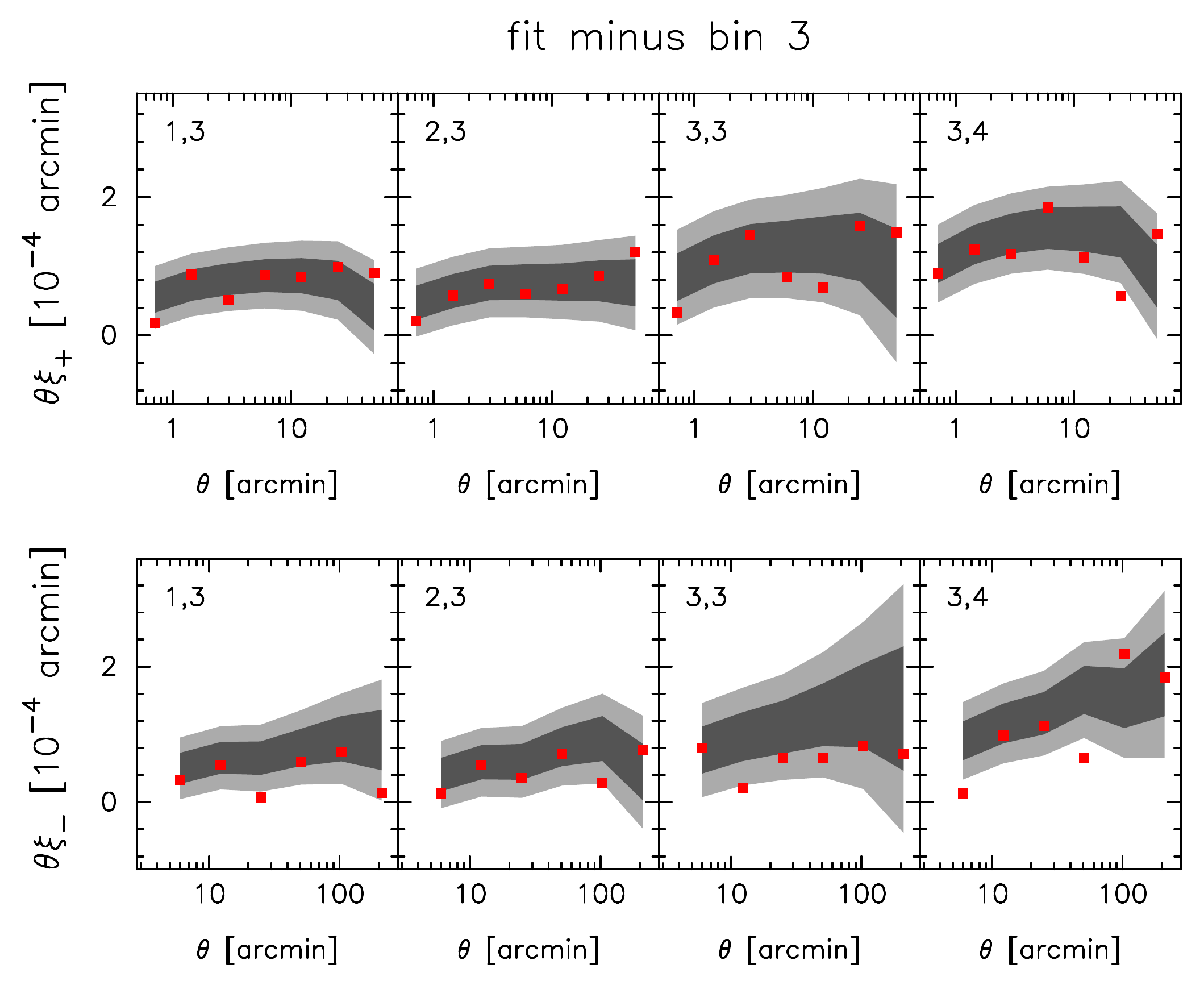} \\
	\includegraphics[width=85mm, angle=0]{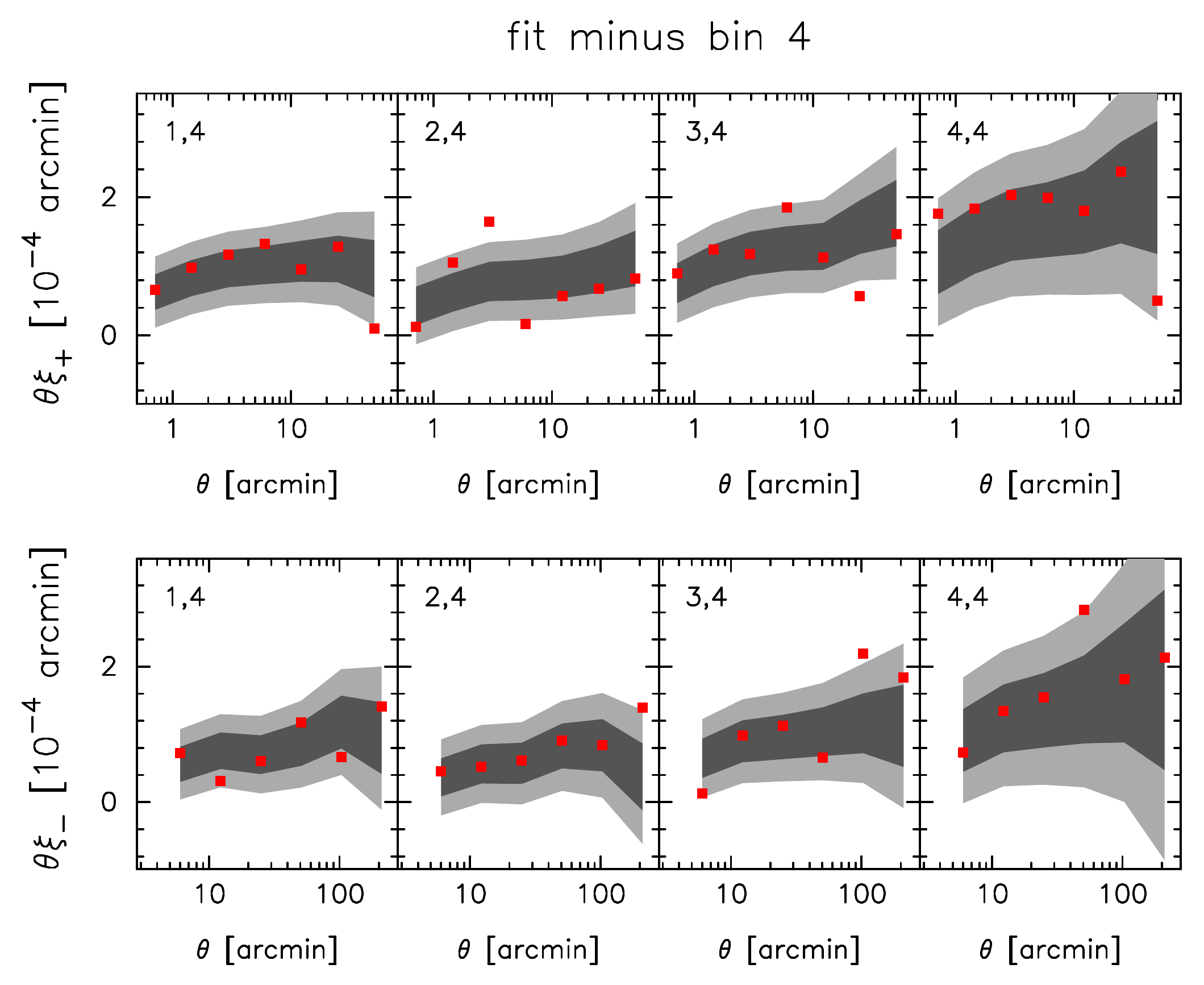}
\vskip -0.07 truein
	\caption{The upper two panels show cross-correlations  $\xi_+$ and $\xi_-$
involving tomographic redshift bin 3 (red points). The numbers in each plot identify the
cross-correlation (e.g. $1, 3$ denotes redshift bin 1 crossed with redshift bin 3). The grey bands
show the allowed $\pm 1 \sigma$ (dark grey) and $\pm 2 \sigma$ (light grey) ranges allowed
by the fits to the rest of the data. The lower two panels show the equivalent plots, but for
cross-correlations involving tomographic redshift bin 4.}
	\label{fig:conditionals}

\end{figure}

Instead of using an amplitude parameter $\lambda$, we can and make a prediction for the
vector ${\bf x}^D$ conditional on the fit to ${\bf y}^D$
\begin{equation}
        {\bf x}^{\rm cond} =  {\hat {\bf x}} + {\bf C}_{xy} {\bf C}_{yy}^{-1} ({\bf y}^D - {\hat {\bf y}}).   \label{cond1}
\end{equation}
If the best-fit model is known exactly, the  covariance of ${\bf x}^{\rm cond}$ is
\begin{equation}
 {\bf C}_{xx}^{\rm cond} = {\bf C}_{xx} - {\bf C}_{xy} {\bf C}_{yy}^{-1} {\bf C}_{yx}.  \label{cond2}
\label{equ:conditional}
\end{equation}
However, in our application the best-fit model is determined by fitting the data vector ${\bf y}^D$ and so 
the uncertainty in the best-fit model contributes an additional variance to ${\bf C}_{xx}^{\rm cond}$:
\begin{equation}
{\bf C^\prime}_{xx}^{\rm cond} = {\bf C}_{xx}^{\rm cond} +  {\bf \Delta C}_{xx}^{\rm cond}, \label{cond2b}
\end{equation}
which we determine empirically by sampling over the MCMC chains. In our application, ${\bf \Delta C}_{xx}^{\rm cond}$ is a small
correction to ${\bf C}_{xx}^{\rm cond}$.

As a test of the consistency of the data we compute  a conditional $\chi^2$:
\begin{equation}
 \chi^2_{\rm cond} = ({\bf x}^D - {\bf x}^{\rm cond})^T({\bf C^\prime}_{xx}^{\rm cond})^{-1} ({\bf x}^D - {\bf x}^{\rm cond}).  \label{cond3}
\end{equation}
The results of these tests are summarized in Table \ref{tab:kids_cond_all} and are consistent with the $\lambda$-tests shown in Fig. \ref{fig:lambda}. Eliminating  $\xi_-$ leads to a substantial increase in $S_8$
that is incompatible with $\xi_-$  at about  $2.7\sigma$. The redshift bin 3 component of the data vector is inconsistent with the rest of the data vector at about $2.6 \sigma$. However, the $\chi^2_{\rm cond}$ reveals a new inconsistency: the
redshift bin 4 component of the data vector is inconsistent with the rest of the data vector at about $3.5 \sigma$.

The origin of the high values of  $\chi^2_{\rm cond}$ for these various partitions of the data vector 
is clear from Fig. \ref{fig:conditionals}. The figure shows the data vector (red points) for all cross-correlations
involving redshift bin 3 (upper two panels) and those involving redshift bin 4 (lower two panels)
compared to the expectations ${\bf x}^{\rm cond}$ conditional on the rest of the data
(equ. \ref{cond1}). The grey bands show $\pm 1$ and $\pm 2 \sigma$ ranges around
${\bf x}^{\rm cond}$ computed from the diagonal components of  equ. \ref{cond2b}.  The top two panels of Fig. \ref{fig:conditionals}
show that cross-correlations involving redshift bin 3 want a lower amplitude than the rest of the data.
This problem is particularly acute for $\xi_{-}$ for the $(3, 3)$ and $(3,4)$ redshift bin cross-correlations.
These two cross-correlations carry quite high weight in fits to the full data vector (driving $S_8$ down),
yet they are inconsistent at nearly $\sim 2.6 \sigma$ with the rest of the data. A possible explanation for this
discrepancy is an inaccuracy in the calibration of the photometric redshifts for bin 3. In fact
\cite{vanUitert:17} present evidence for a $2.3\sigma$ negative shift of $\Delta z \approx -0.06$ for this redshift bin.
They find no evidence for significant shifts in the other redshift bins.

As summarized in Table \ref{tab:kids_cond_all}, removing redshift bin
4 lowers the value of $S_8$ but increases the errors on $S_8$
substantially because the intrinsic alignment amplitude is less well
constrained. From Fig. \ref{fig:conditionals} this low amplitude
solution appears to match reasonably well with the general shape of
the rest of the data vector, but now we see a high
value of $\chi^2_{\rm cond}$ arising from {\it outliers}. In the lower
two panels of this figure, 8 out of 52 data points sit outside the
conditional $\pm 2 \sigma$ range\footnote{Assuming Gaussian
  statistics, the $p-$value for this is about $2.4 \times
  10^{-3}$.}. Several of these outliers are at large angular scales
and are not obvious in plots using errors computed from the diagonals
of the full covariance matrix (e.g. Fig. 5 of H17). However, the KiDS
covariance matrix tells us that the data vector should be correlated
across different tomographic redshift bins.  What
Fig. \ref{fig:conditionals} shows is that the KiDS correlation
functions display significantly higher variance than expected from the
KiDS covariance matrix, particularly at large angular scales and for
correlations involving redshift bin 4. This excess variance is a
serious problem because it means that the KiDS errors on cosmological
parameters are systematically underestimated, especially if data at
small angular scales is excluded.

Our analysis shows strong evidence for a statistical inconsistency between
the KiDS estimates of $\xi_+$ and $\xi_-$.
H17 and \cite{vanUitert:17} find evidence for non-zero B-modes in the
KiDS data at small angular scales ($\theta < 4.2^\prime$), indicative
of systematics. If  systematic errors contribute equally to the
tangential and cross distortions (and this has not been demonstrated
for KiDS), then the B-modes will affect $\xi_+$, but not
$\xi_-$. Eliminating $\xi_+$ entirely from the fits lowers $S_8$ to
$0.705 \pm 0.048$ (see Table \ref{tab:kids_cond_all}) with
$\chi^2=82.2$ for $50$ degrees of freedom (a $3.2\sigma$ excess). In other 
words, if one argues that the difference between $\xi_+$ and $\xi_-$ is
indicative of  systematic errors in $\xi_+$, then the tension between KiDS and  
\Planck\ is  exacerbated.

\section{Comparison with other techniques for measuring the amplitude of the fluctuation spectrum}

The results of the previous section show that there are some worrying
internal inconsistencies in the KiDS dataset as analysed in H17. These
inconsistencies suggest that we should be cautious in interpreting the
KiDS constraints on cosmology.  However, the tests in themselves do
not tell us the causes of the inconsistencies, or their impact on the
estimates of $S_8$.  Is the amplitude of the matter fluctuations at
redshifts $z \simlt 1$ really lower than expected in the
\Plancks\ \LCDM\ cosmology? 

\begin{figure}
	\centering
	\includegraphics[width=82mm, angle=0]{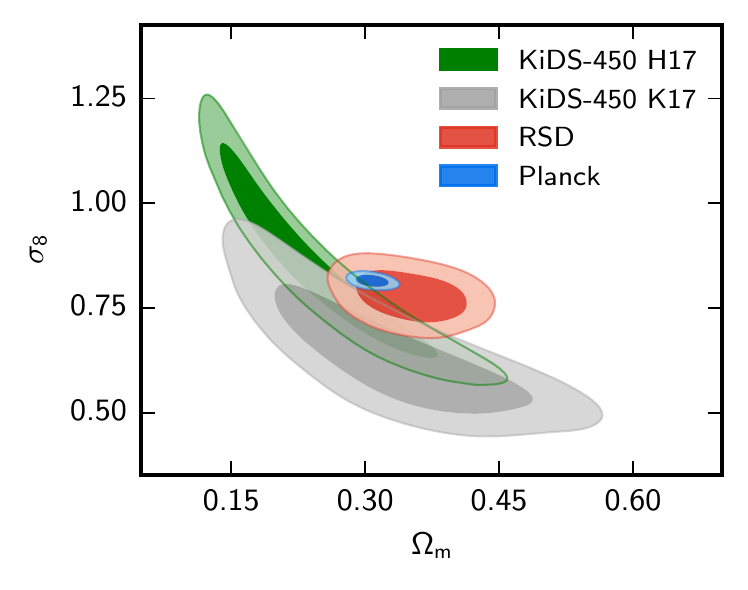} 
	\caption{Constraints in the $\sigma_8-\Omega_m$ plane assuming the spatially flat \LCDM\ cosmology. The 68\% and 95\% contours from \Plancks are shown in blue. The constraints from the H17 fiducial KiDS analysis are shown in green. The grey contours show the constraints from the power-spectrum analysis of KiDS reported by K17. The red contours show the constraints
from redshift-space distortions (RSD) as discussed in the text.}
	\label{fig:summary}
\end{figure}

Another way of studying the amplitude of the matter power spectrum is via redshift
space distortions \citep[RSD, ][]{Kaiser:87}. RSD provide a measurement of the parameter
combination $f\sigma_8$, where $f$ is the logarithmic derivative of the linear growth rate
with respect to the scale factor 
\begin{equation}
f = {{ \rm d \ ln} D \over { \rm d \ ln} a}, 
\end{equation}
and $a = (1+z)^{-1}$. In the \LCDM\ model, $f \approx \Omega_m(z)^{0.55}$
and so RSD measure the parameter combination $\sigma_8
\Omega_m^{0.55}$, {\it i.e.} similar to the  parameter
combination $S_8$ up to a known constant. Measurements of RSD from
the DR12 analysis of the Baryon Oscillation Spectroscopy Survey (BOSS)
have been reported by \cite{Alam:16}. These measurements are for three
redshift slices with effective redshifts $z_{\rm eff}=0.38$, $z_{\rm
  eff}=0.51$ and $z_{\rm eff}=0.61$, substantially overlapping with
the redshift range of the KiDS survey.  \cite{Huterer:17} have
recently used the Supercal Type Ia supernova compilation
\citep{Scolnic:15} together with independent distance measurements of
galaxies \citep{Springob:14} to measure $f\sigma_8$ at $z_{\rm eff}=0.02$.
The \Plancks \LCDM\ cosmology is in excellent agreement with these
measurements of $f\sigma_8$ over the entire redshift range  $z =
0.02-0.61$. The consistency between \Plancks and the RSD measurements
is illustrated in Fig. \ref{fig:summary}, where we have combined the
BOSS and Supercal RSD measurements to produce constraints in the
$\sigma_8- \Omega_m$ plane\footnote{This is done using the
  final$\_$consensus$\_$dV$\_$FAP$\_$fsig data files and covariance
  matrix downloaded from
  https://sdss3.org/science/boss$\_$publications.php. We then scanned
  the likelihood, using uniform priors in $H_0$ and $\Omega_m h^2$ to
  rescale the BOSS distance $D_V$ and Alcock-Paczynski
  \citep{Alcock:79} parameter $F_{AP}$ to the fiducial sound horizon
  used in the BOSS analysis, fixing $\Omega_bh^2$ to the P16
  \LCDM\ value.}. The RSD constraints are in mild tension with the
KiDS correlation function analysis of H17, and in even greater tension
with the tomographic power-spectrum analysis of KiDS described by
K17  {\it using the same shear catalogue}.

The abundance of rich clusters of galaxies (selected at various wavelengths) has
been used in a number of studies to constrain the amplitude of the fluctuations
spectrum at low redshift \citep[e.g.][]{Vikhlinin:09, Rozo:10, Hasselfield:13, 
Planck_clusters:14, Mantz:15, Planck_clusters:16, deHaan:16}. As summarized in several
of these papers, calibration of cluster masses is a major source of uncertainty in this type of analysis. Two recent studies \citep{ Mantz:15, deHaan:16} use weak gravitational lensing mass estimates from the `Weighing the Giants' programme \citep{vonderLinden:14, Kelly:14,Applegate:14} to calibrate cluster
scaling relations. \cite{Mantz:15} use an X-ray selected sample of clusters from the ROSAT
All-Sky Survey covering the redshift range $0<z<0.5$, finding $\sigma_8(\Omega_m/0.3)^{0.17} 
= 0.81\pm 0.03$. \cite{deHaan:16} use a sample of clusters identified with the South Pole Telescope
with median redshift $z_{\rm med} = 0.53$ to infer $\sigma_8(\Omega_m/0.27)^{0.3} = 0.797 \pm 0.031$.
Both of these estimates are consistent with the Planck P16 \LCDM\ cosmology: $\sigma_8(\Omega_m/0.3)^{0.17} = 0.818 \pm 0.009$, $\sigma_8(\Omega_m/0.27)^{0.3} = 0.848 \pm 0.012$. Thus, there is no convincing evidence for any discrepancy between rich cluster counts and the expectations from the \Planck-\LCDM\ cosmology. The \cite{deHaan:16} study is particularly interesting because it
covers a similar redshift range to those of the BOSS RSD and KiDS measurements, yet is consistent with \Planck\ and RSD.

\section{Comparison of weak lensing estimates of ${\bf S_8}$: the importance of intrinsic alignments}

Fig. \ref{fig:summary} shows a discrepancy between the H17 and K17 analyses, which are based on the same shear catalogue. There is little doubt that the H17 and K17
analyses are incompatible,  since not one of the 
$14,469$ samples in the  K17 MCMC likelihood chain\footnote{KiDS450$\_$QE$\_$EB$\_$4bins$\_$3zbins$\_$basez$\_$ia$\_$bary$\_$nu.txt,
  downloaded from http://kids.strw.leidenuniv.nl.} has  parameters
close to those of the best fit found by H17. In fact, \cite{vanUitert:17} (hereafter vU17) have
computed cross power-spectra from $\xi_+$ and $\xi_-$ for the KiDS
data using the identical redshift bins to those used in K17. Their
auto-spectrum for the highest redshift bin differs substantially from
the quadratic estimate of K17. The origin of this difference is not
understood\footnote{Note that the quadratic estimator used by K17 is
  sensitive to noise estimation, particularly if there are B-mode
  systematics (which are known to be present in the KiDS
  data). Inaccurate noise estimation would primarily affect the
  auto-spectra, where the noise levels are high compared to the
  cosmological signal (see Fig. 4 of H17).}. Another pointer that the K17
results are affected by systematic errors comes from the intrinsic 
alignment solution.  K17 find $A_{IA} = -1.72^{+1.49}_{-1.25}$ which has the
opposite (and from the theoretical perspective, counterintuitive) sign
to that found by H17. This difference drives down the amplitude of
$S_8$ in the K17 analysis. Both the direct comparison of spectra reported
by \cite{vanUitert:17} and the shift to a negative intrinsic alignment 
amplitude suggest that the K17 analysis is suspect.

The key point that we want to emphasise here is that the intrinsic
alignment parameter $A_{IA}$ is not a benign `nuisance' parameter \citep[for reviews see e.g.][]{Troxel:14, Joachimi:15} The modelling of
intrinsic alignments 
is degenerate with the cosmological parameters of interest,
$\sigma_8$, $\Omega_m$, and $S_8$, and so the model and associated parameters matter. 
Systematic errors in the data can be absorbed by the intrinsic alignment model
 and this will have an impact on cosmology.
For example,  \cite{vanUitert:17} have noted that the parameter $A_{IA}$ can absorb systematic
errors in the calibrations of photometric redshift distributions. (This can also be inferred from
Fig. \ref{fig:ia+s8} which shows the sensitivity of the intrinsic alignment solution for the KiDS
data to the highest photometric redshift bin). Implausible
(e.g. strongly negative) values of $A_{IA}$ suggest systematic errors and should therefore
be followed up.

\begin{figure}
	\centering
	\includegraphics[width=85mm, angle=0]{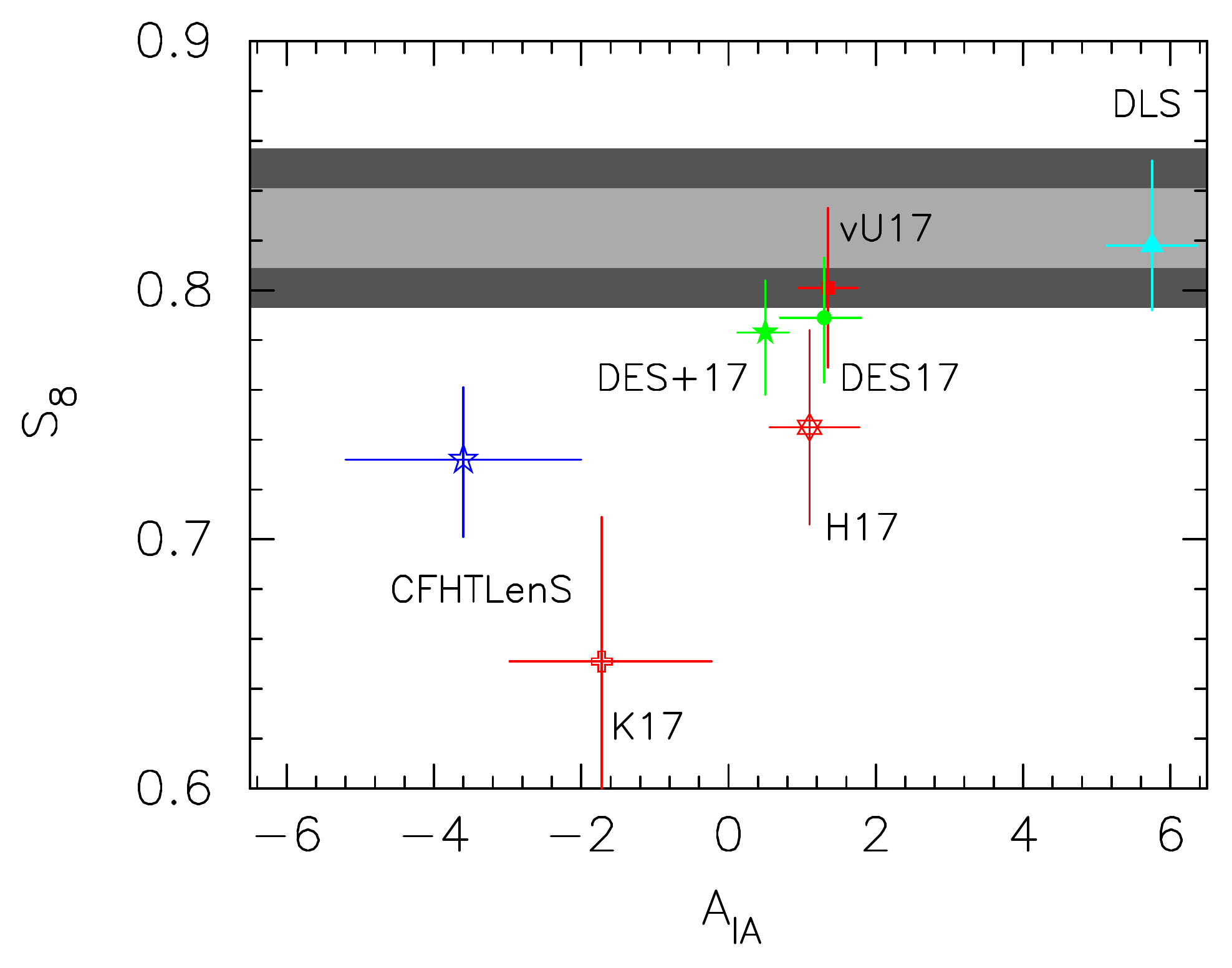}	
	\caption{$S_8$ plotted against the intrinsic alignment amplitude for various surveys together with $1\sigma$ errors on $S_8$ and $A_{IA}$.
The grey bands show the $1\sigma$ and $2\sigma$ constraints from \Planck. The data points are
as follows: CFHTLens \citep{Joudaki:17a}; DLS \citep{Jee:16}\protect\footnotemark; K17
shows the power spectrum analysis of KiDS \citep{Kohlinger:17}; H17 shows the correlation function analysis of KiDS \citep{Hildebrandt:17}; vU17 shows the constraints from combining $P^{gg}$, $P^{gm}$ and $P^E$ measurements from KiDS and GAMA data \citep{vanUitert:17}; DES17 shows the cosmic shear constraints from DES  year 1 data \citep{Troxel:17} (note that the DES analyses uses a
redshift dependent model of intrinsic aligments, as described in the text); DES+17 shows the combination of DES year 1 cosmic shear results with galaxy-galaxy and galaxy-shear measurements \citep{DES:17}. }

	\label{fig:IA_v_S8}

\end{figure}

\footnotetext{Note that the
\cite{Jee:16} `baseline' analysis of DLS use a luminosity dependent model
of intrinsic alignments and impose a flat prior of $5.14 < A_{IA} <
6.36$, motivated by the results of \cite{Joachimi:11}. However, they find that
their results on $S_8$ are insensitive to $A_{IA}$ (see their Fig. 12),
presumably because of the huge depth of DLS.}

As an example, one of the lowest weak lensing determinations of $S_8$ 
comes from the reanalysis of the revised CFHTLenS data
\citep{Joudaki:16}. However, these authors  find a strongly
 negative value of $A_{IA} = -3.6 \pm 1.6$, a value which seems unlikely for any reasonable mix
of galaxy types. The recent DES analysis of
\cite{Troxel:17} uses a redshift dependent amplitude: $A_{IA} [(1+z)/(1.62)]^{\eta}$,
finding $A_{IA} = 1.3^{+0.5}_{-0.6}$, $\eta=3.7^{+1.0}_{-2.3}$ \footnote{These constraints become
 $A_{IA} = 0.5^{+0.32}_{-0.38}$, $\eta = 0^{+2.7}_{-2.8}$ with the addition of 
galaxy-galaxy and galaxy-shear data, \cite{DES:17}). These authors argue that an amplitude
of $A_{IA} \sim 0.5$ is consistent with their selection criteria if only red galaxies contribute
to the intrinsic alignments.}. \cite{Troxel:17} also test a more elaborate `mixed' alignment
model based on the work of \cite{Blazek:17}. This model leads to a downward shift of $S_8$ by about $1\sigma$, demonstrating that uncertainties in the modelling of intrinsic alignments 
makes a non-negligible contribution to the errors in cosmological parameters.

Returning to the KiDS survey, one way of achieving better control of intrinsic alignments 
and photometric redshift calibration errors is to
add additional types of data. vU17 have analysed
the shear power spectra from KiDS, $P^{E}$ (constructed by integrating
over $\xi_+$ and $\xi_-$). In addition, they use the Galaxies Mass Assembly (GAMA)
redshift survey \citep{Driver:11, Liske:15} to compute the galaxy-mass power-spectra, $P^{gm}$
by cross-correlating the KiDS shear measurements with GAMA galaxies, and the galaxy-galaxy power
spectra $P^{gg}$. From $P^{gm}+P^{gg}$, they find $S_8 = 0.853 \pm 0.042$. Combining with $P^E$, they find $A_{IA} = 1.30 \pm 0.40$ and  $S_8 = 0.801 \pm 0.032$ (consistent with the \Planck\ and RSD results shown in Fig. \ref{fig:summary}). 



Figure \ref{fig:IA_v_S8} gives a summary of the results discussed in this Section. The two analyses
that are most discrepant with the $S_8$ value from Planck (CFHTLenS and K17) both have 
strongly negative intrinsic alignment solutions. The H17 results are in tension with \Planck\
but become consistent with \Planck\ with the addition of galaxy-galaxy and galaxy-mass data
(vU17). The DES year 1 analyses plotted in Fig. \ref{fig:IA_v_S8} 
are both consistent with \Planck. The intrinsic alignment solutions of vU17 and \cite{DES:17}
(i.e. $A_{IA} \sim  1$)
seem physically plausible given the mix of galaxy types expected in these surveys.

\section{Conclusions}

The main purpose of this paper has been to highlight and quantify internal inconsistencies in 
the KiDS cosmic shear analysis. 
Our main conclusion is that more effort is needed to resolve
inconsistencies in the KiDS data. This includes understanding the
origin of the B-modes, systematic differences between $\xi_+$ and $\xi_-$, the
parameter shifts seen by excluding photometric redshift bin 3, the
large excess $\chi^2$ and  scatter at large angular scales.  Until this is
done, it seems premature to draw inferences on new physics from KiDS.

Comparison of \Planck\ with other measures of the amplitude of the mass
fluctuations, principally redshift space distortions from BOSS, reveals
no evidence for any inconsistencies with the \Planck\ base-\LCDM\ cosmology.
We have also reviewed cosmic shear constraints on $S_8$, emphasising the 
degeneracy between intrinsic alignments and cosmology. As summarized in Fig. \ref{fig:IA_v_S8}
the two analyses which yield the lowest values of $S_8$ both have strongly negative values of
$A_{IA}$. The DES 1 year analyses are consistent with the \Planck\ \LCDM\ value for $S_8$
\citep{Troxel:17, DES:17} and give physically plausive values for $A_{IA}$. The H17 value of
$S_8$ from KiDS sits about $2.3 \sigma$ low compared to \Planck, but is pulled upwards
with the addition of galaxy-galaxy, galaxy-mass data (vU17). Overall, we conclude 
there is no strong evidence for any inconsistency between the \Planck\ \LCDM\ cosmology
and  measures of the amplitude of the fluctuation spectrum at low redshift.

\section*{Acknowledgements}
We thank Hiranya Peiris, Benjamin Joachimi, Fergus Simpson  and the referee for
helpful comments on the preprint version of this paper.
We thank Frankie Nobis-Efstathiou for help with the early stages of
this project. We also thank Anthony Challinor, Steven Gratton and
members of the KiDS team for comments on aspects of this analysis.
We also thank members of the Planck Parameters team.
Pablo Lemos acknowledges support from an Isaac Newton Studentship
at the University of Cambridge and from the Science and Technologies
Facilities Council.

\bibliographystyle{mnras}
\bibliography{kids_pablo} 

\begin{thebibliography}{}
\makeatletter
\relax
\def\mn@urlcharsother{\let\do\@makeother \do\$\do\&\do\#\do\^\do\_\do\%\do\~}
\def\mn@doi{\begingroup\mn@urlcharsother \@ifnextchar [ {\mn@doi@}
  {\mn@doi@[]}}
\def\mn@doi@[#1]#2{\def\@tempa{#1}\ifx\@tempa\@empty \href
  {http://dx.doi.org/#2} {doi:#2}\else \href {http://dx.doi.org/#2} {#1}\fi
  \endgroup}
\def\mn@eprint#1#2{\mn@eprint@#1:#2::\@nil}
\def\mn@eprint@arXiv#1{\href {http://arxiv.org/abs/#1} {{\tt arXiv:#1}}}
\def\mn@eprint@dblp#1{\href {http://dblp.uni-trier.de/rec/bibtex/#1.xml}
  {dblp:#1}}
\def\mn@eprint@#1:#2:#3:#4\@nil{\def\@tempa {#1}\def\@tempb {#2}\def\@tempc
  {#3}\ifx \@tempc \@empty \let \@tempc \@tempb \let \@tempb \@tempa \fi \ifx
  \@tempb \@empty \def\@tempb {arXiv}\fi \@ifundefined
  {mn@eprint@\@tempb}{\@tempb:\@tempc}{\expandafter \expandafter \csname
  mn@eprint@\@tempb\endcsname \expandafter{\@tempc}}}

\bibitem[\protect\citeauthoryear{{Abbott} et~al.,}{{Abbott}
  et~al.}{2016}]{Abbott:16}
{Abbott} T.,  et~al., 2016, \mn@doi [\prd] {10.1103/PhysRevD.94.022001}, 94,
  022001

\bibitem[\protect\citeauthoryear{{Alam} et~al.,}{{Alam} et~al.}{2016}]{Alam:16}
{Alam} S.,  et~al., 2016, preprint (\mn@eprint {arXiv} {1607.03155})

\bibitem[\protect\citeauthoryear{{Alcock} \& {Paczynski}}{{Alcock} \&
  {Paczynski}}{1979}]{Alcock:79}
{Alcock} C.,  {Paczynski} B.,  1979, \mn@doi [\nat] {10.1038/281358a0}, 281,
  358

\bibitem[\protect\citeauthoryear{{Amendola} et~al.,}{{Amendola}
  et~al.}{2016}]{Amendola:16}
{Amendola} L.,  et~al., 2016, preprint (\mn@eprint {arXiv} {1606.00180})

\bibitem[\protect\citeauthoryear{{Applegate} et~al.,}{{Applegate}
  et~al.}{2014}]{Applegate:14}
{Applegate} D.~E.,  et~al., 2014, \mn@doi [\mnras] {10.1093/mnras/stt2129},
  \href {http://adsabs.harvard.edu/abs/2014MNRAS.439...48A} {439, 48}

\bibitem[\protect\citeauthoryear{{Blandford}, {Saust}, {Brainerd}  \&
  {Villumsen}}{{Blandford} et~al.}{1991}]{Blandford:91}
{Blandford} R.~D.,  {Saust} A.~B.,  {Brainerd} T.~G.,   {Villumsen} J.~V.,
  1991, \mn@doi [\mnras] {10.1093/mnras/251.4.600}, 251, 600

\bibitem[\protect\citeauthoryear{{Blazek}, {MacCrann}, {Troxel}  \&
  {Fang}}{{Blazek} et~al.}{2017}]{Blazek:17}
{Blazek} J.,  {MacCrann} N.,  {Troxel} M.~A.,   {Fang} X.,  2017, preprint,
  \href {http://adsabs.harvard.edu/abs/2017arXiv170809247B} {} (\mn@eprint
  {arXiv} {1708.09247})

\bibitem[\protect\citeauthoryear{{Bridle} \& {King}}{{Bridle} \&
  {King}}{2007}]{Bridle:07}
{Bridle} S.,  {King} L.,  2007, \mn@doi [New Journal of Physics]
  {10.1088/1367-2630/9/12/444}, 9, 444

\bibitem[\protect\citeauthoryear{{DES Collaboration} et~al.,}{{DES
  Collaboration} et~al.}{2017}]{DES:17}
{DES Collaboration} et~al., 2017, preprint, \href
  {http://adsabs.harvard.edu/abs/2017arXiv170801530D} {} (\mn@eprint {arXiv}
  {1708.01530})

\bibitem[\protect\citeauthoryear{{Driver} et~al.,}{{Driver}
  et~al.}{2011}]{Driver:11}
{Driver} S.~P.,  et~al., 2011, \mn@doi [\mnras]
  {10.1111/j.1365-2966.2010.18188.x}, 413, 971

\bibitem[\protect\citeauthoryear{{Hasselfield} et~al.,}{{Hasselfield}
  et~al.}{2013}]{Hasselfield:13}
{Hasselfield} M.,  et~al., 2013, \mn@doi [\jcap]
  {10.1088/1475-7516/2013/07/008}, \href
  {http://adsabs.harvard.edu/abs/2013JCAP...07..008H} {7, 008}

\bibitem[\protect\citeauthoryear{{Heymans} et~al.,}{{Heymans}
  et~al.}{2012}]{Heymans:12}
{Heymans} C.,  et~al., 2012, \mn@doi [\mnras]
  {10.1111/j.1365-2966.2012.21952.x}, 427, 146

\bibitem[\protect\citeauthoryear{{Heymans} et~al.,}{{Heymans}
  et~al.}{2013}]{Heymans:13}
{Heymans} C.,  et~al., 2013, \mn@doi [\mnras] {10.1093/mnras/stt601}, 432, 2433

\bibitem[\protect\citeauthoryear{{Hildebrandt} et~al.,}{{Hildebrandt}
  et~al.}{2017}]{Hildebrandt:17}
{Hildebrandt} H.,  et~al., 2017, \mn@doi [\mnras] {10.1093/mnras/stw2805}, 465,
  1454

\bibitem[\protect\citeauthoryear{{Hinshaw} et~al.,}{{Hinshaw}
  et~al.}{2013}]{WMAP:13}
{Hinshaw} G.,  et~al., 2013, \mn@doi [\apjs] {10.1088/0067-0049/208/2/19}, 208,
  19

\bibitem[\protect\citeauthoryear{{Hirata} \& {Seljak}}{{Hirata} \&
  {Seljak}}{2004}]{Hirata:04}
{Hirata} C.~M.,  {Seljak} U.,  2004, \mn@doi [\prd]
  {10.1103/PhysRevD.70.063526}, 70, 063526

\bibitem[\protect\citeauthoryear{{Huterer}, {Shafer}, {Scolnic}  \&
  {Schmidt}}{{Huterer} et~al.}{2017}]{Huterer:17}
{Huterer} D.,  {Shafer} D.~L.,  {Scolnic} D.~M.,   {Schmidt} F.,  2017, \mn@doi
  [\jcap] {10.1088/1475-7516/2017/05/015}, 5, 015

\bibitem[\protect\citeauthoryear{{Jee}, {Tyson}, {Hilbert}, {Schneider},
  {Schmidt}  \& {Wittman}}{{Jee} et~al.}{2016}]{Jee:16}
{Jee} M.~J.,  {Tyson} J.~A.,  {Hilbert} S.,  {Schneider} M.~D.,  {Schmidt} S.,
   {Wittman} D.,  2016, \mn@doi [\apj] {10.3847/0004-637X/824/2/77}, 824, 77

\bibitem[\protect\citeauthoryear{{Joachimi}, {Mandelbaum}, {Abdalla}  \&
  {Bridle}}{{Joachimi} et~al.}{2011}]{Joachimi:11}
{Joachimi} B.,  {Mandelbaum} R.,  {Abdalla} F.~B.,   {Bridle} S.~L.,  2011,
  \mn@doi [\aap] {10.1051/0004-6361/201015621}, \href
  {http://adsabs.harvard.edu/abs/2011A%26A...527A..26J} {527, A26}

\bibitem[\protect\citeauthoryear{{Joachimi} et~al.,}{{Joachimi}
  et~al.}{2015}]{Joachimi:15}
{Joachimi} B.,  et~al., 2015, \mn@doi [\ssr] {10.1007/s11214-015-0177-4}, \href
  {http://adsabs.harvard.edu/abs/2015SSRv..193....1J} {193, 1}

\bibitem[\protect\citeauthoryear{{Joudaki} et~al.,}{{Joudaki}
  et~al.}{2016}]{Joudaki:16}
{Joudaki} S.,  et~al., 2016, preprint (\mn@eprint {arXiv} {1610.04606})

\bibitem[\protect\citeauthoryear{{Joudaki} et~al.,}{{Joudaki}
  et~al.}{2017}]{Joudaki:17a}
{Joudaki} S.,  et~al., 2017, \mn@doi [\mnras] {10.1093/mnras/stw2665}, 465,
  2033

\bibitem[\protect\citeauthoryear{{Kaiser}}{{Kaiser}}{1987}]{Kaiser:87}
{Kaiser} N.,  1987, \mn@doi [\mnras] {10.1093/mnras/227.1.1}, 227, 1

\bibitem[\protect\citeauthoryear{{Kaiser}}{{Kaiser}}{1992}]{Kaiser:92}
{Kaiser} N.,  1992, \mn@doi [\apj] {10.1086/171151}, 388, 272

\bibitem[\protect\citeauthoryear{{Kelly} et~al.,}{{Kelly}
  et~al.}{2014}]{Kelly:14}
{Kelly} P.~L.,  et~al., 2014, \mn@doi [\mnras] {10.1093/mnras/stt1946}, \href
  {http://adsabs.harvard.edu/abs/2014MNRAS.439...28K} {439, 28}

\bibitem[\protect\citeauthoryear{{Kirk}, {Rassat}, {Host}  \& {Bridle}}{{Kirk}
  et~al.}{2012}]{Kirk:12}
{Kirk} D.,  {Rassat} A.,  {Host} O.,   {Bridle} S.,  2012, \mn@doi [\mnras]
  {10.1111/j.1365-2966.2012.21099.x}, 424, 1647

\bibitem[\protect\citeauthoryear{{K{\"o}hlinger} et~al.,}{{K{\"o}hlinger}
  et~al.}{2017}]{Kohlinger:17}
{K{\"o}hlinger} F.,  et~al., 2017, preprint (\mn@eprint {arXiv} {1706.02892})

\bibitem[\protect\citeauthoryear{{Liske} et~al.,}{{Liske}
  et~al.}{2015}]{Liske:15}
{Liske} J.,  et~al., 2015, \mn@doi [\mnras] {10.1093/mnras/stv1436}, 452, 2087

\bibitem[\protect\citeauthoryear{{Mantz} et~al.,}{{Mantz}
  et~al.}{2015}]{Mantz:15}
{Mantz} A.~B.,  et~al., 2015, \mn@doi [\mnras] {10.1093/mnras/stu2096}, \href
  {http://adsabs.harvard.edu/abs/2015MNRAS.446.2205M} {446, 2205}

\bibitem[\protect\citeauthoryear{{Miralda-Escude}}{{Miralda-Escude}}{1991}]{Miralda-Escude:91}
{Miralda-Escude} J.,  1991, \mn@doi [\apj] {10.1086/170555}, 380, 1

\bibitem[\protect\citeauthoryear{{Planck Collaboration} et~al.,}{{Planck
  Collaboration} et~al.}{2014a}]{Planckparams14}
{Planck Collaboration} et~al., 2014a, \mn@doi [\aap]
  {10.1051/0004-6361/201321591}, 571, A16

\bibitem[\protect\citeauthoryear{{Planck Collaboration} et~al.,}{{Planck
  Collaboration} et~al.}{2014b}]{Planck_clusters:14}
{Planck Collaboration} et~al., 2014b, \mn@doi [\aap]
  {10.1051/0004-6361/201321521}, \href
  {http://adsabs.harvard.edu/abs/2014A%26A...571A..20P} {571, A20}

\bibitem[\protect\citeauthoryear{{Planck Collaboration} et~al.,}{{Planck
  Collaboration} et~al.}{2016a}]{Planckparams16}
{Planck Collaboration} et~al., 2016a, \mn@doi [\aap]
  {10.1051/0004-6361/201525830}, 594, A13

\bibitem[\protect\citeauthoryear{{Planck Collaboration} et~al.,}{{Planck
  Collaboration} et~al.}{2016b}]{Plancklensing16}
{Planck Collaboration} et~al., 2016b, \mn@doi [\aap]
  {10.1051/0004-6361/201525941}, 594, A15

\bibitem[\protect\citeauthoryear{{Planck Collaboration} et~al.,}{{Planck
  Collaboration} et~al.}{2016c}]{Planck_clusters:16}
{Planck Collaboration} et~al., 2016c, \mn@doi [\aap]
  {10.1051/0004-6361/201525833}, \href
  {http://adsabs.harvard.edu/abs/2016A%26A...594A..24P} {594, A24}

\bibitem[\protect\citeauthoryear{{Rozo} et~al.,}{{Rozo} et~al.}{2010}]{Rozo:10}
{Rozo} E.,  et~al., 2010, \mn@doi [\apj] {10.1088/0004-637X/708/1/645}, \href
  {http://adsabs.harvard.edu/abs/2010ApJ...708..645R} {708, 645}

\bibitem[\protect\citeauthoryear{{Scolnic} et~al.,}{{Scolnic}
  et~al.}{2015}]{Scolnic:15}
{Scolnic} D.,  et~al., 2015, \mn@doi [\apj] {10.1088/0004-637X/815/2/117},
  \href {http://ukads.nottingham.ac.uk/abs/2015ApJ...815..117S} {815, 117}

\bibitem[\protect\citeauthoryear{{Sievers} et~al.,}{{Sievers}
  et~al.}{2013}]{ACT13}
{Sievers} J.~L.,  et~al., 2013, \mn@doi [\jcap]
  {10.1088/1475-7516/2013/10/060}, 10, 060

\bibitem[\protect\citeauthoryear{{Springob} et~al.,}{{Springob}
  et~al.}{2014}]{Springob:14}
{Springob} C.~M.,  et~al., 2014, \mn@doi [\mnras] {10.1093/mnras/stu1743},
  \href {http://ukads.nottingham.ac.uk/abs/2014MNRAS.445.2677S} {445, 2677}

\bibitem[\protect\citeauthoryear{{Story} et~al.,}{{Story} et~al.}{2013}]{SPT13}
{Story} K.~T.,  et~al., 2013, \mn@doi [\apj] {10.1088/0004-637X/779/1/86}, 779,
  86

\bibitem[\protect\citeauthoryear{{Troxel} \& {Ishak}}{{Troxel} \&
  {Ishak}}{2015}]{Troxel:14}
{Troxel} M.~A.,  {Ishak} M.,  2015, \mn@doi [\physrep]
  {10.1016/j.physrep.2014.11.001}, \href
  {http://adsabs.harvard.edu/abs/2015PhR...558....1T} {558, 1}

\bibitem[\protect\citeauthoryear{{Troxel} et~al.,}{{Troxel}
  et~al.}{2017}]{Troxel:17}
{Troxel} M.~A.,  et~al., 2017, preprint, \href
  {http://adsabs.harvard.edu/abs/2017arXiv170801538T} {} (\mn@eprint {arXiv}
  {1708.01538})

\bibitem[\protect\citeauthoryear{{Vikhlinin} et~al.,}{{Vikhlinin}
  et~al.}{2009}]{Vikhlinin:09}
{Vikhlinin} A.,  et~al., 2009, \mn@doi [\apj] {10.1088/0004-637X/692/2/1060},
  \href {http://adsabs.harvard.edu/abs/2009ApJ...692.1060V} {692, 1060}

\bibitem[\protect\citeauthoryear{{de Haan} et~al.,}{{de Haan}
  et~al.}{2016}]{deHaan:16}
{de Haan} T.,  et~al., 2016, \mn@doi [\apj] {10.3847/0004-637X/832/1/95}, \href
  {http://adsabs.harvard.edu/abs/2016ApJ...832...95D} {832, 95}

\bibitem[\protect\citeauthoryear{{van Uitert} et~al.,}{{van Uitert}
  et~al.}{2017}]{vanUitert:17}
{van Uitert} E.,  et~al., 2017, preprint (\mn@eprint {arXiv} {1706.05004})

\bibitem[\protect\citeauthoryear{{von der Linden} et~al.,}{{von der Linden}
  et~al.}{2014}]{vonderLinden:14}
{von der Linden} A.,  et~al., 2014, \mn@doi [\mnras] {10.1093/mnras/stt1945},
  \href {http://adsabs.harvard.edu/abs/2014MNRAS.439....2V} {439, 2}

\makeatother
\end{thebibliography}
\end{document}